\documentclass[aps,prb,superscriptaddress,amsmath,amssymb,floats,twocolumn,longbibliography,floatfix
]{revtex4-1}

\usepackage{graphicx}
\usepackage{placeins}
\usepackage{dcolumn}
\usepackage{bm}
\usepackage{color}

\begin{document}

\preprint{APS/123-QED}

\title{Nature of the field-induced magnetic incommensurability in multiferroic Ni$_3$TeO$_6$}

\author{J. Lass}
\affiliation{Nanoscience Center, Niels Bohr Institute, University of Copenhagen, DK-2100 Copenhagen \O , Denmark}
\affiliation{Laboratory for Neutron Scattering and Imaging, Paul Scherrer Institute, CH-5232 Villigen, Switzerland}

\author{Ch. R\o hl Andersen} 
\affiliation{Nanoscience Center, Niels Bohr Institute, University of Copenhagen, DK-2100 Copenhagen \O , Denmark}
\affiliation{National Centre for Nano Fabrication and Characterization, Technical University of Denmark, 2800 Kgs. Lyngby, Denmark}

\author{H. K. Leerberg} 
\author{S. Birkemose} 
\affiliation{Nanoscience Center, Niels Bohr Institute, University of Copenhagen, DK-2100 Copenhagen \O , Denmark}
 
\author{S. Toth} 
\author{U. Stuhr} 
\author{M. Bartkowiak} 
\author{Ch. Niedermayer} 
\affiliation{Laboratory for Neutron Scattering and Imaging, Paul Scherrer Institute, CH-5232 Villigen, Switzerland}

\author{Zhilun Lu} 
\affiliation{Helmholtz-Zentrum Berlin, D-14109 Berlin Wannsee, Germany}

\author{R. Toft-Petersen} 
\affiliation{Institute of Physics, Technical University of Denmark, DK-2800 Lyngby, Denmark}

\author{M. Retuerto} 
\affiliation{Instituto de Catálisis y Petroleoquímica, Consejo Superior de Investigaciones Cientificas, Cantoblanco E-28049, Madrid, Spain}

\author{J. Okkels Birk} 
\author{K. Lefmann} 
\affiliation{Nanoscience Center, Niels Bohr Institute, University of Copenhagen, DK-2100 Copenhagen \O , Denmark}


\date{\today}

\begin{abstract}
Using single crystal neutron diffraction we show that the magnetic structure Ni$_3$TeO$_6$ at fields above 8.6~T along the $c$ axis and low temperature changes from a commensurate collinear antiferromagnetic structure with spins along $c$ and ordering vector $Q_{\rm C} = (0 \; 0 \; 1.5)$, to a conical spiral with propagation vector 
$Q_{\rm IC} = (0 \; 0 \; 1.5 \pm \delta)$, $\delta \sim 0.18$, having a significant spin component in the $(a,b)$ plane. We determine the phase diagram of this material in magnetic fields up to 10.5~T along $c$ and show the phase transition between the low field and conical spiral phases is of first order by observing a discontinuous jump of the ordering vector. $Q_{\rm IC}$ is found to drift both as function of magnetic field and temperature.
Preliminary inelastic neutron scattering data reveals that the spin wave gap in zero field has minima exactly at $Q_{\rm IC}$ and a gap of about 1.1~meV consisting with a cross-over around ~8.6 T. Further, a simple magnetic Hamiltonian accounting in broad terms for these is presented. 
Our findings confirms the exclusion of the inverse Dzyaloshinskii-Moriya interaction as a cause for the giant magneto-electric due to symmetry arguments. In its place we advocate for the symmetric exchange striction as the origin of this effect.
\end{abstract}

\pacs{Valid PACS appear here}

\maketitle


\section{Introduction}
Multiferroic materials display an intriguing coupling between structural, magnetic and electronic order~\cite{Spalding2005,Cheong2007}. 
These properties make multiferroics especially interesting for applications in multi-functional devices, e.g. in spintronics and as transducers, actuators, capacitors, sensors, or multi-memory devices~\cite{hill2000,Lee2008a,Catalan2009}. 
One particular property searched for is the control of magnetic order by an applied electrical field. This effect is controlled through the magneto-electric (M-E) coupling and with a large coupling strength one would be able to easily change polarisation, which in turn is a great leap forward for magnetic data storage technology~\cite{Chun2012,Ryan2013}.

The family of hexagonal tellurides M$_3$TeO$_6$ (M being a transition metal) represents a popular class of multiferroics\cite{Hudl2011,Mathieu2013,Harris2012,Wang2013,Her2011,Ivanov2012,Li2012,Choi2008}. A strong interest has arisen in nickel telluride, Ni$_3$TeO$_6$ (NTO), as it displays a giant M-E coupling close to a field-induced magnetic phase transition at $\sim$8.6~T along the $c$-axis at low temperature\cite{ONeal2018,Zivkovic2010,Oh2014,Yokosuk2015,Yokosuk2016,Zivkovic2010,Skiadopoulou2017,Raman2013}. In particular, it was found that the system switches between a spin and electric polarized commensurate (C) state to an incommensurate spin state with lower electric polarization with hardly any hysteresis, below 1 mT - a property that could ultimately lead to loss-free magnetoelectric devices~\cite{Oh2014}.

These observations have been attributed to a continuous spin-flop transition between two antiferromagnetic phases, through a narrow intermediate phase~\cite{Oh2014}. 
In the same work, a significant M-E effect was observed in the region around the phase transition - one of the largest M-E effects observed in any single-phase material~\cite{Yokosuk2015}. 
A later study showed an even stronger M-E effect taking place at a second field-induced phase transition at 52~T~\cite{Kim2015}.

In this work, we show that the magnetic phase transition at $\sim$8.6~T takes place from the C state to an incommensurate spiral spin structure (IC) and is in fact a first-order transition, which is at variance with views in present literature~\cite{Oh2014,Yokosuk2016}.
We find that the width of the phase transition i.e. the co-existence region is of the order 0.4~T and we further map out the phase diagram for temperatures below 60~K and magnetic fields below 10.5~T. 
A preliminary inelastic study provides insight into the magnetic couplings and allow for the creation of a simplified magnetic Hamiltonian describing the phase transition and the low energy excitations. Our findings have important consequences for the understanding of the origin of the magnetoelectric effect in NTO, where the possibility of an inverse  Dzyaloshinskii-Moriya(DM)-driven phase transition earlier was excluded due to Landau theory\cite{Oh2014}. We can exclude it using model-free arguments on the basis of the relation between the ordering vector and electric polarization. We support the suggestion of the symmetric exchange striction to change the pyroelectric low-field phase into a paraelectric high-field phase.

\section{Sample growth and characterization}
%

\begin{figure}[ht]
\includegraphics[height=3.7cm]{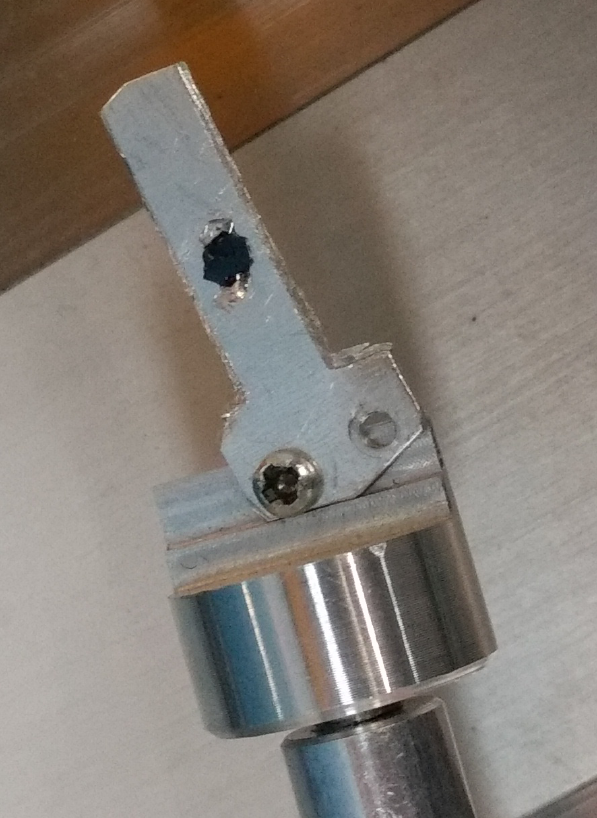} \vspace{2mm}\includegraphics[height=3.7cm]{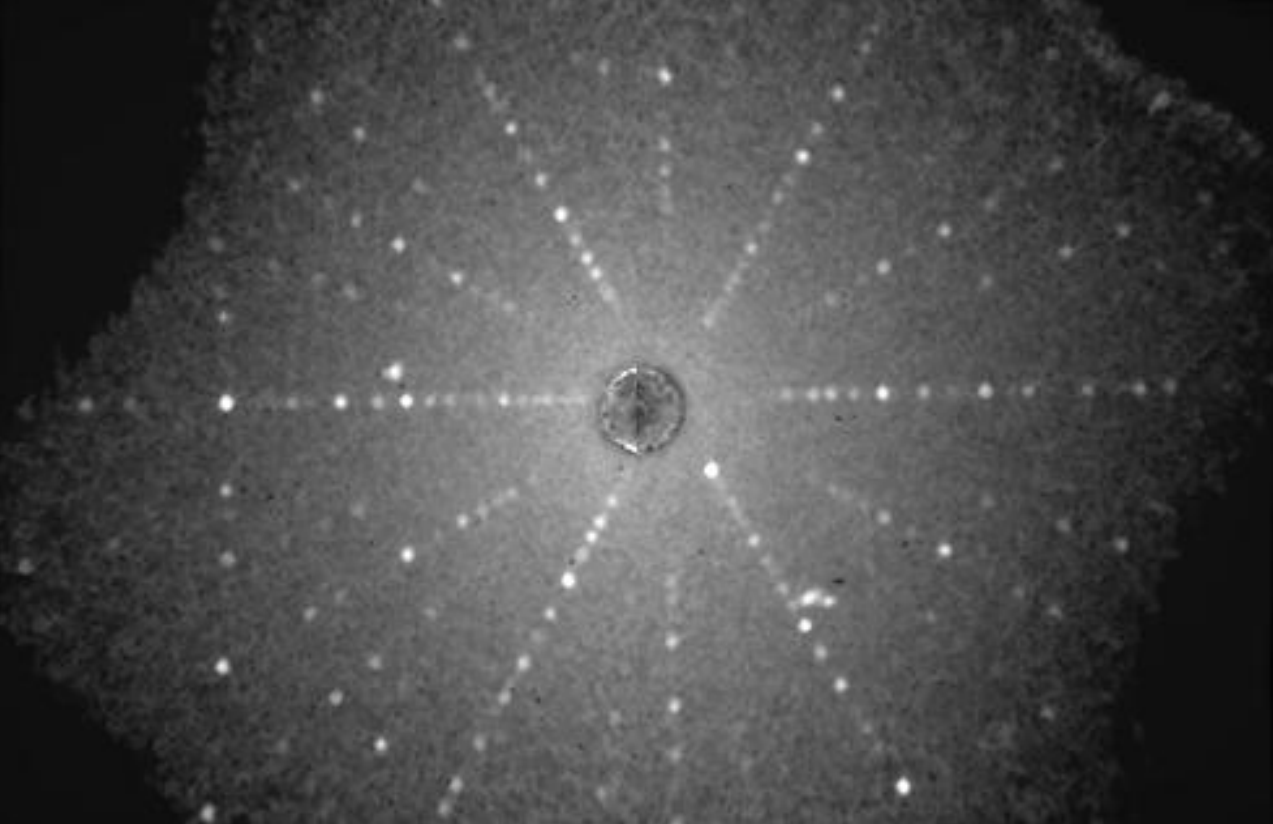}
\caption{(left) NTO single crystal sample of $\approx$ 15~mg used in diffraction experiments. (right) White-beam X-ray Laue pattern obtained on this crystal showing its good quality.} \label{fig:XRLaue}
\end{figure}

Powdered NTO (previously prepared), V$_2$O$_5$, TeO$_2$, NaCl and KCl in a molar ratio of 1:5:10:10:5 were mixed and placed in an alumina crucible.  The mixture was heated at 830$^\circ$C for 3 days and then slowly cooled down to 600$^\circ$C during five days. 
The resulting batch of single crystals were small platelets of typical sizes $4 \times 4 \times 0.5$~mm$^3$ and with masses of 5-15~mg. They were tested with backscattering white-beam X-ray Laue diffraction using an Ag anode. A photograph of the diffraction crystal is shown in Fig.~\ref{fig:XRLaue} along with its Laue pattern. The sharpness of the Laue peaks is a signature of a low intrinsic mosaicity (below $2^\circ)$ of the crystal. Later neutron diffraction measurements proved the mosaicity to be below 0.3$^\circ$.

For the neutron diffraction experiments one single crystal of mass 15~mg was selected, while for the inelastic neutron scattering, we used a mosaic of 12 crystals with a total mass of 106.9~mg, co-aligned with Laue X-ray diffraction in the $(a,c)$ plane and checked by diffration to be within $2^\circ$. 

\begin{figure}[ht]\centering
\includegraphics[trim={0cm 0cm 0cm 0cm},clip,height=5.5cm]{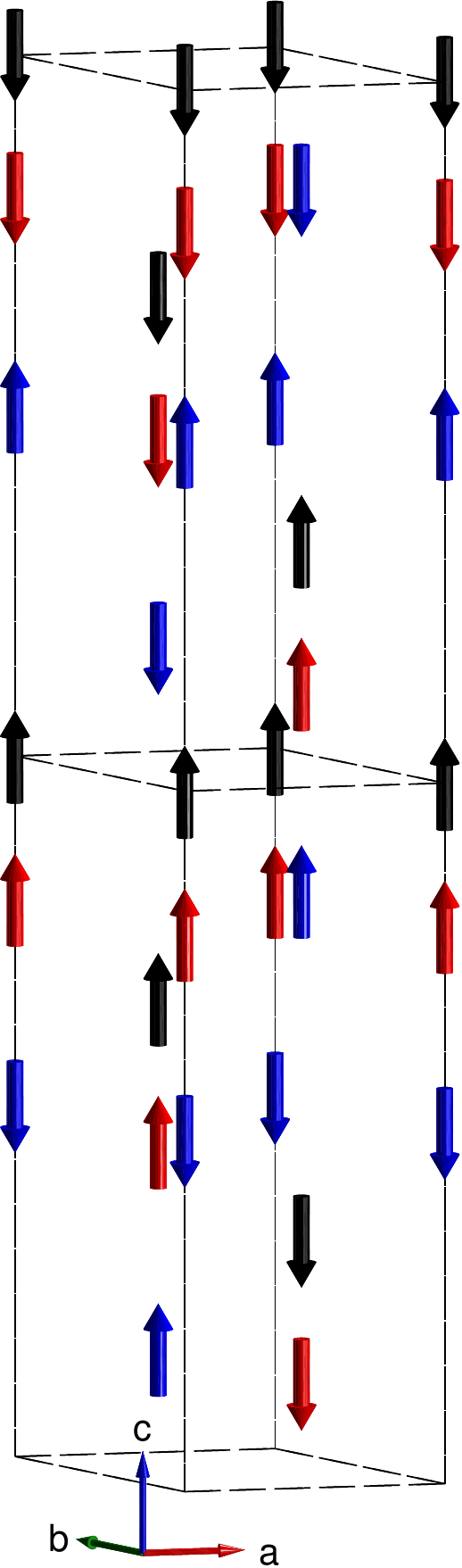}\hspace{0.5cm}
\includegraphics[trim={0cm 0cm 0cm 0cm},clip,height=5.5cm]{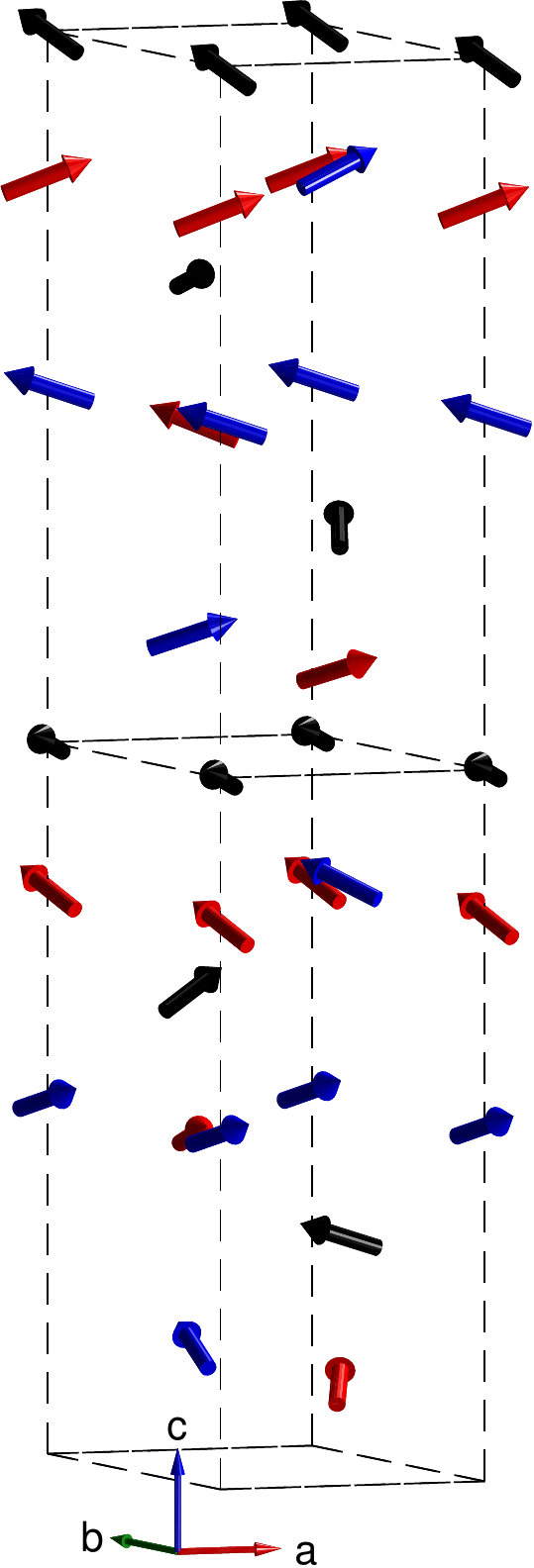}\hspace{0.5cm}
\includegraphics[height=5.775cm]{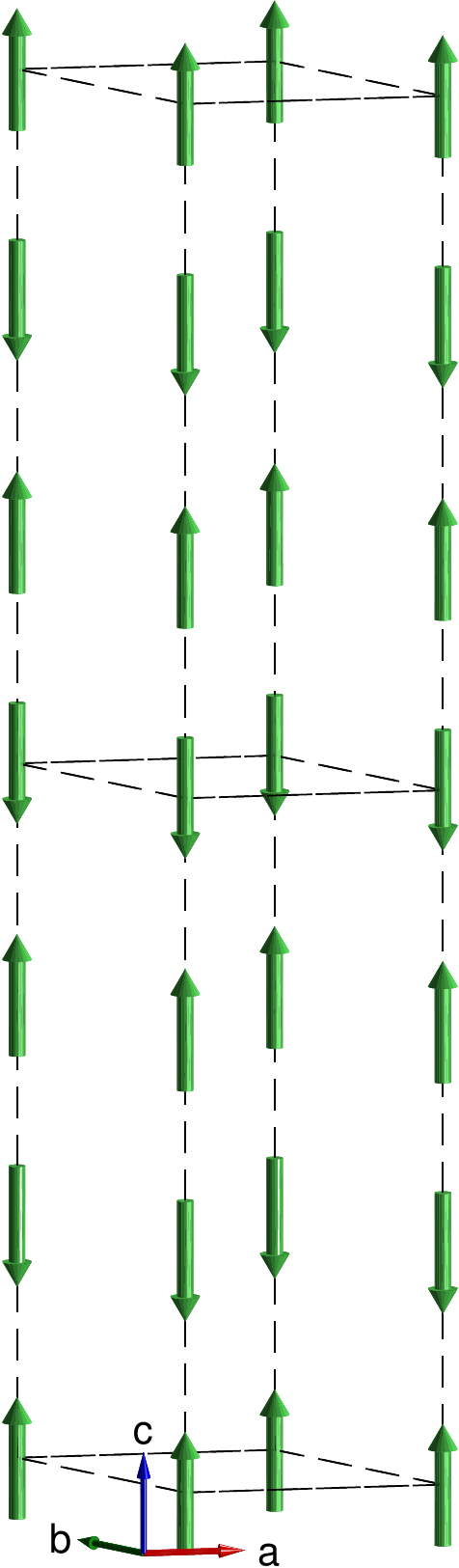}
\caption{
(left) Model of the zero-field magnetic structure of NTO~\cite{Zivkovic2010}. (middle) Our suggestion for high-field magnetic structure of NTO. (right) Simplified model used to emulate excitations and phase transition. Green horizontal lines denote $J_{\rm FM}$ while vertical lines are a combination of $J_1$ and $J_2$ coupling nearest and next nearest neighbours respectively.}
\label{fig:magneticstructure}
\end{figure}

\section{Neutron scattering experiments}
\begin{figure}[ht]\centering
\includegraphics[width=\linewidth]{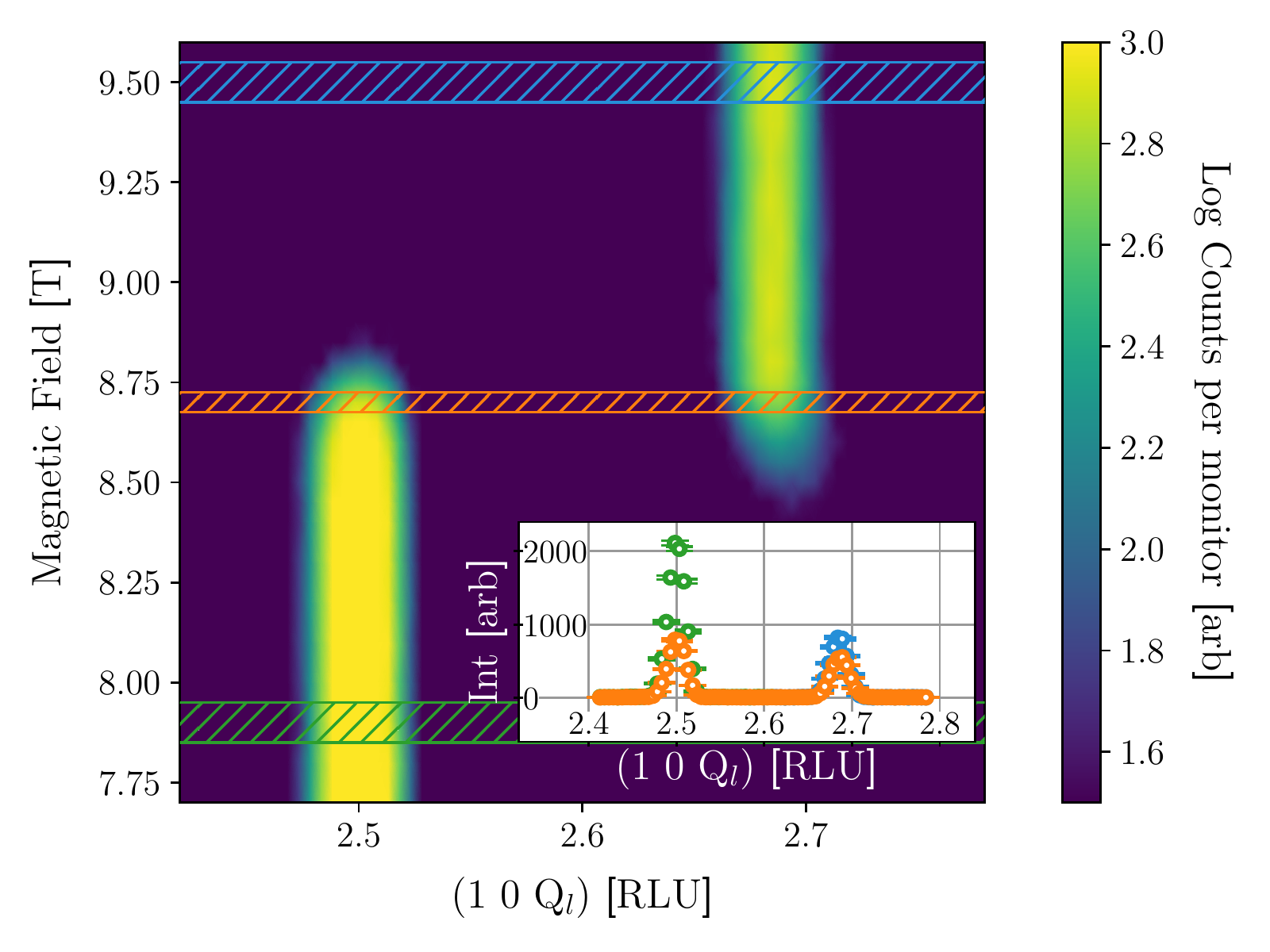}

\includegraphics[width=\linewidth]{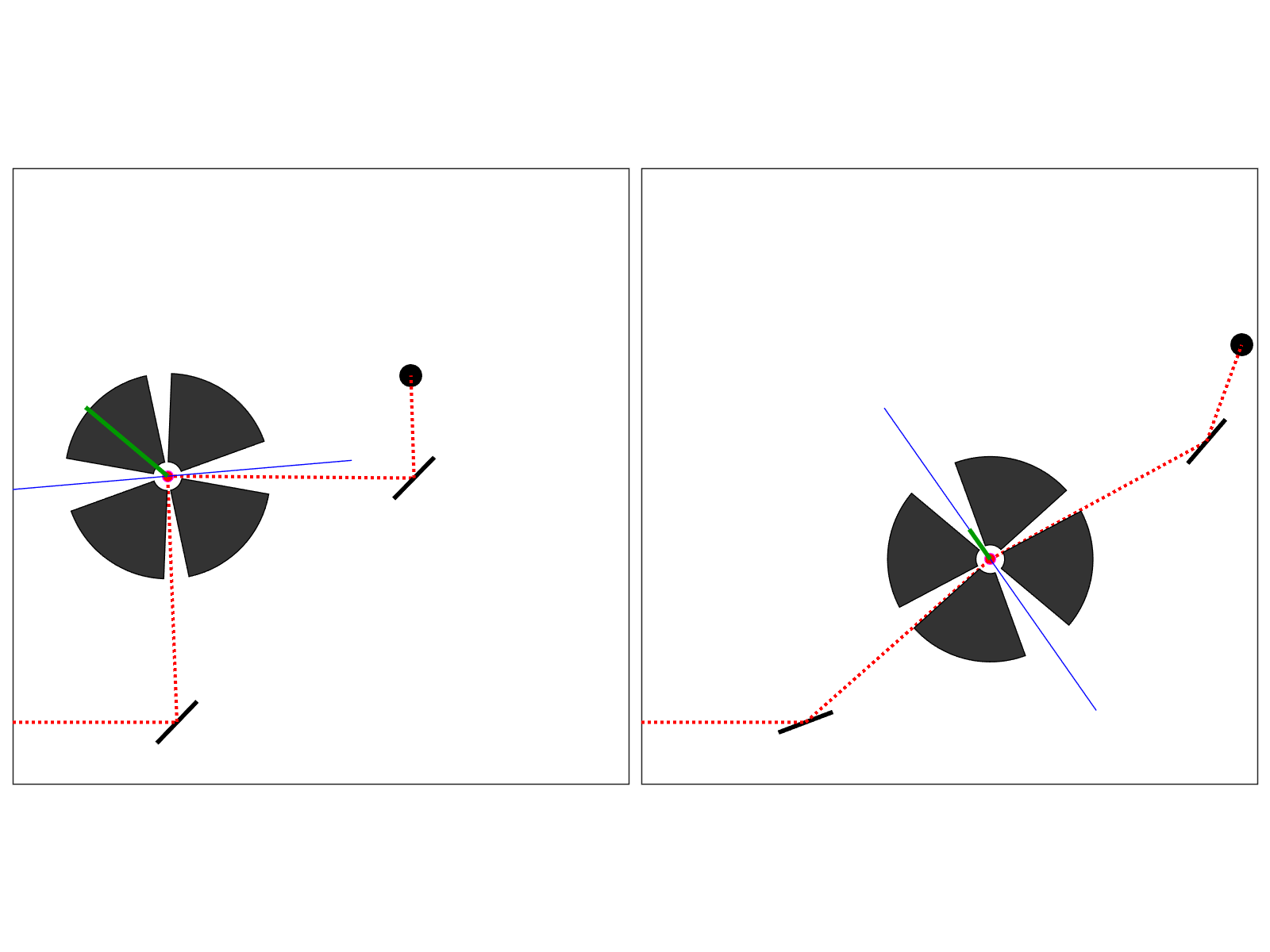}
\caption{(top) 
Magnetic diffraction intensity at $(1 \; 0 \; l)$ taken at RITA-II at 1.7~K, shown as a function of $l$ and field. Colors signify the corresponding phase as depicted in the phase diagram in Fig.~\ref{fig:phasediagram}. (insert) Scattering intensities for the three highlighted areas. (bottom) Sketch of the 11~T horizontal-field magnet used for the neutron scattering experiments. The lines show possible neutron beam paths through the one window, scattering off the sample, and exiting again through another window.} \label{fig:rawdataANDmagnetwindows}
\end{figure}

NTO was investigated with neutron diffraction using a horizontal-field cryomagnet at both the cold RITA-II~\cite{bahlrita1,bahlrita2} and the thermal EIGER~\cite{Eiger} triple axis instruments with the sample orientations ($h$\;0\;$l$) and ($h$\;$h$\;$l$) and energies in the range of $E_{\rm i} = E_{\rm f} = 3 - 8$~meV and $E_{\rm i} = E_{\rm f} = 14.7 - 100$~meV, respectively. The applied magnetic fields were up to 10.5 T along $c$ and temperatures between 1.8 K and 200 K. We used effective collimation sequences of open-80'-40'-open and open-80'-80'-open. As the scattering vector lies in the same plane as the field direction a horizontal field cryomagnet was used. It was equipped with four narrow windows, placed $90^{\circ} $ apart, as illustrated in the bottom of Fig.~\ref{fig:rawdataANDmagnetwindows}. The windows allow passage of the neutron beam at scattering angles below $\approx 18^\circ$ and in a window around $90^{\circ} $. As the scattering geometry was quite limited. a full magnetic structure determination was not possible.


Zero-field inelastic neutron scattering was performed at HZB Berlin, using the FLEXX triple-axis instrument~\cite{Le2013} with the MultiFLEXX secondary spectrometer~\cite{groitl17}. Emphasis was put on measuring the lowest lying magnetic excitations at the energy $\sim$1.5~meV in the $(h \; 0 \; l)$ plane.


\section{Results}

\begin{figure}[ht]\centering
\includegraphics[width=\linewidth]{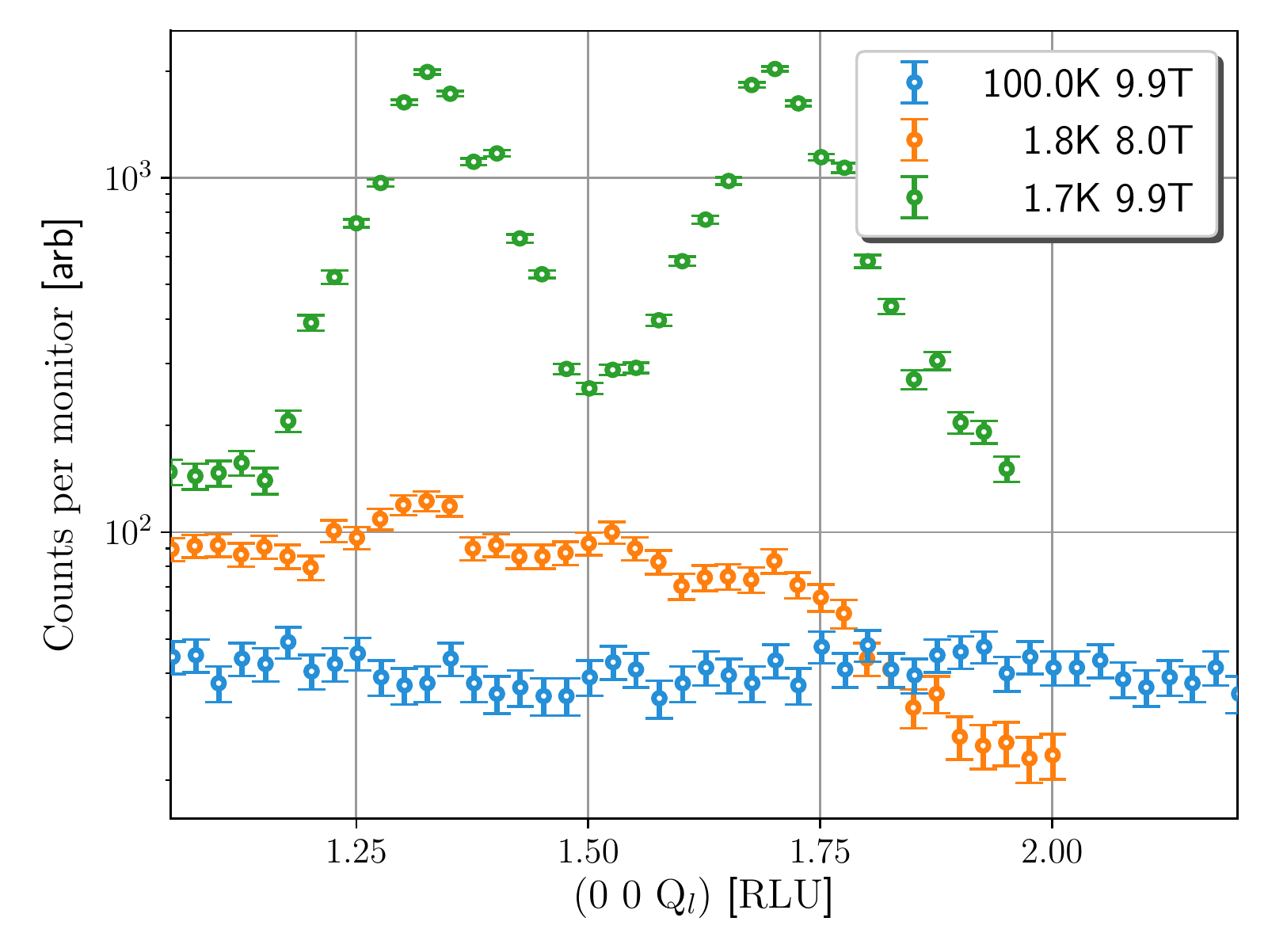}
\includegraphics[width=\linewidth]{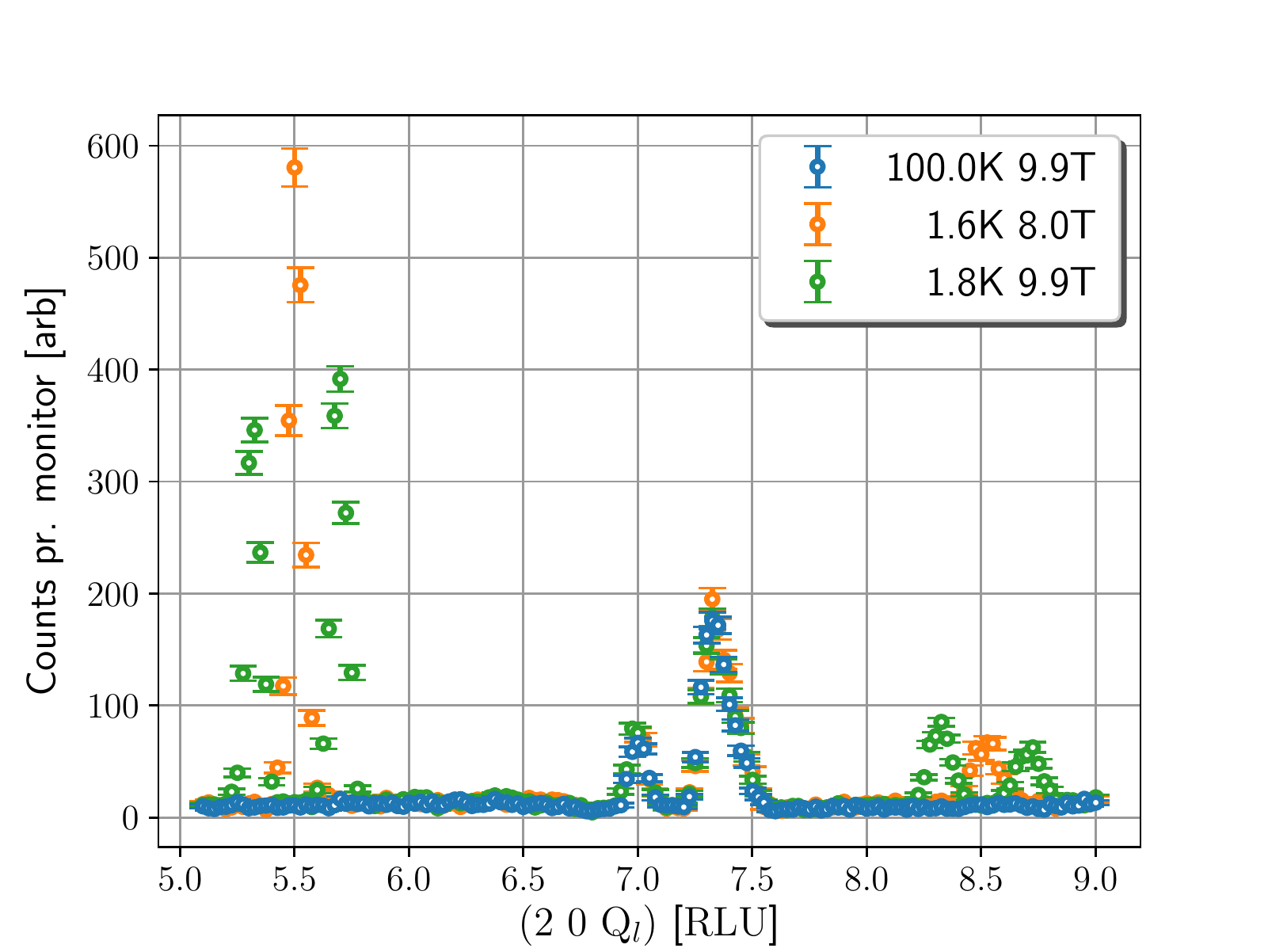}
\caption{Magnetic diffraction data showing scans along (top) $(0 \; 0 \; l)$ and  (bottom)$(2 \; 0 \; l)$ in all three magnetic regions: 1.7~K and 100~K, at $B = 9.9$~T and 2~K, at $B = 8.0$~T. The signal at (2 \; 0 \; 7.3) present across all phases is believed to be spurious of nature.} 
\label{fig:rawdata3}
\end{figure}

We used the good q-resolution of cold neutron diffraction to study the reflection $(1 \; 0\; 2.5 + \delta)$ as function of magnetic field, which is structurally equivalent to (0 \; 0 \; 1.5 + $\delta$). Resulting data 
are shown in the top of Fig.~\ref{fig:rawdataANDmagnetwindows} and in Fig.~\ref{fig:rawdata3}.

In fields up to 8~T, our diffraction data confirm the previously established commensurate antiferromagnetic low-field structure\cite{Zivkovic2010}, illustrated in Fig.~\ref{fig:magneticstructure}. Up to this point, the magnetic Ni$^{2+}$ moments form ferromagnetic layers in the $(a,b)$-plane and are antiferromagnetically ordered along $c$ with a commensurate propagation vector $Q_{\rm C} = (0$\;$0$\;$1.5)$~\cite{Zivkovic2010}. The spin direction is collinear along the $c$-axis, as proven by the complete absence of magnetic intensity from peaks along the $l$-axis, consistent with the selection rules for neutron scattering~\cite{Squires}. There are 6 planes in a magnetic unit cell, corresponding to a doubling of the chemical unit cell along $c$, with alternatingly 1 and 2 Ni atoms per plane within the unit cell, c.f. Fig.~\ref{fig:magneticstructure}. 
We observed a number of magnetic reflections, listed in Tab.~\ref{tab:magneticreflections}, and all observations of peaks and absences are in agreement with this structure. 

Fig.~\ref{fig:rawdata3} shows our findings for higher magnetic field: a consistent split of the commensurate magnetic peaks of the low-field phase into a pair of incommensurate peaks in the high-field phase. The fundamental magnetic ordering vectors are ${\bf Q}_{\rm IC} = (0 \; 0 \; 1.5 \pm \delta)$ with $\delta = 0.18$ for the IC phase and $\delta=0$ for the C phase. 

The field-induced splitting of the commensurate antiferromagnetic peak points to an incommensurate modulation of the ferromagnetic planes along the $l$-direction. The simultaneous appearance of the magnetic peaks along the fundamental $(0 \; 0 \; l)$-direction shows that the spin direction is no longer confined to point along the $c$-axis.



\begin{figure}[ht]\centering
\includegraphics[width=\linewidth]{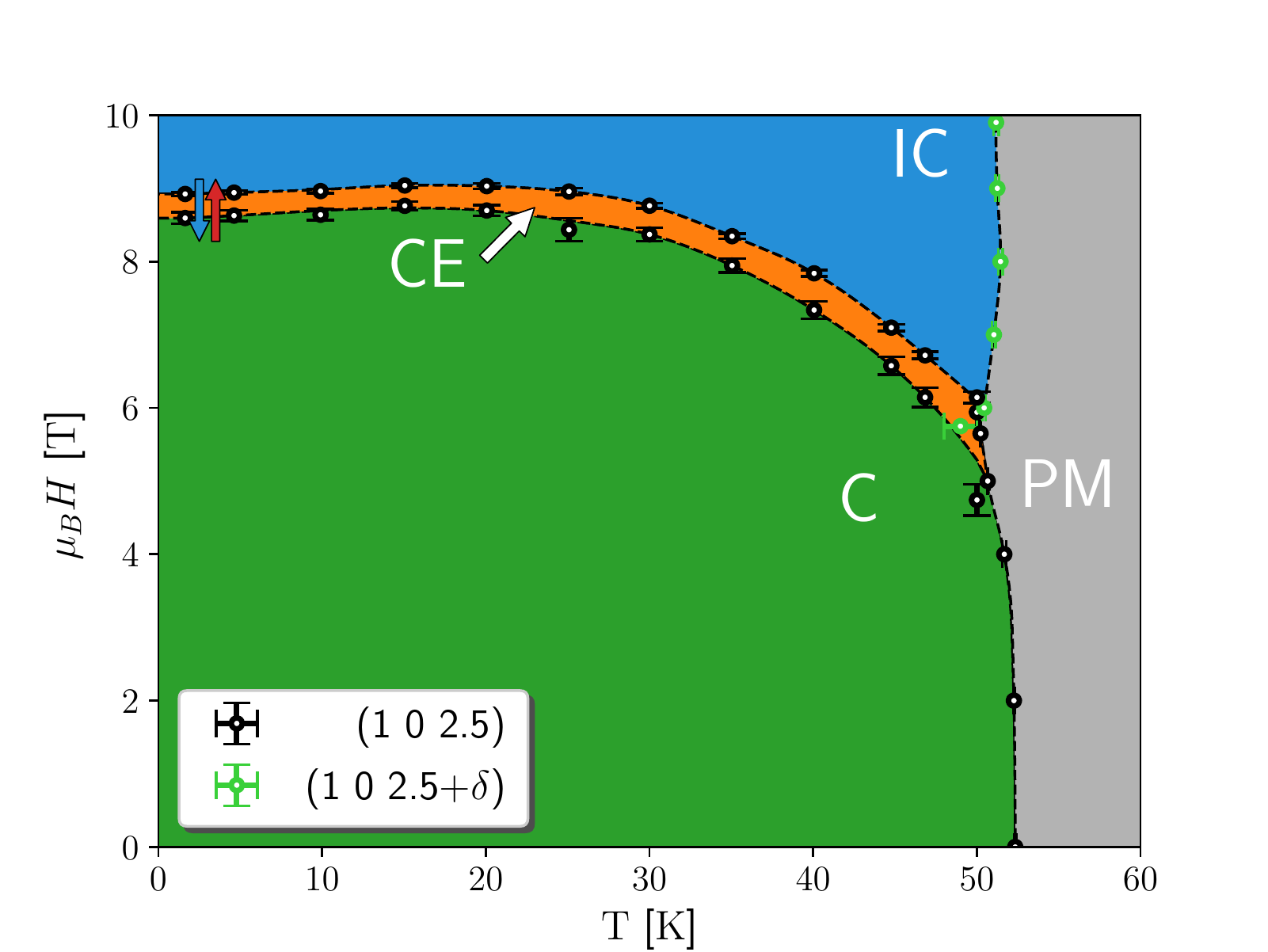}
\includegraphics[width=\linewidth]{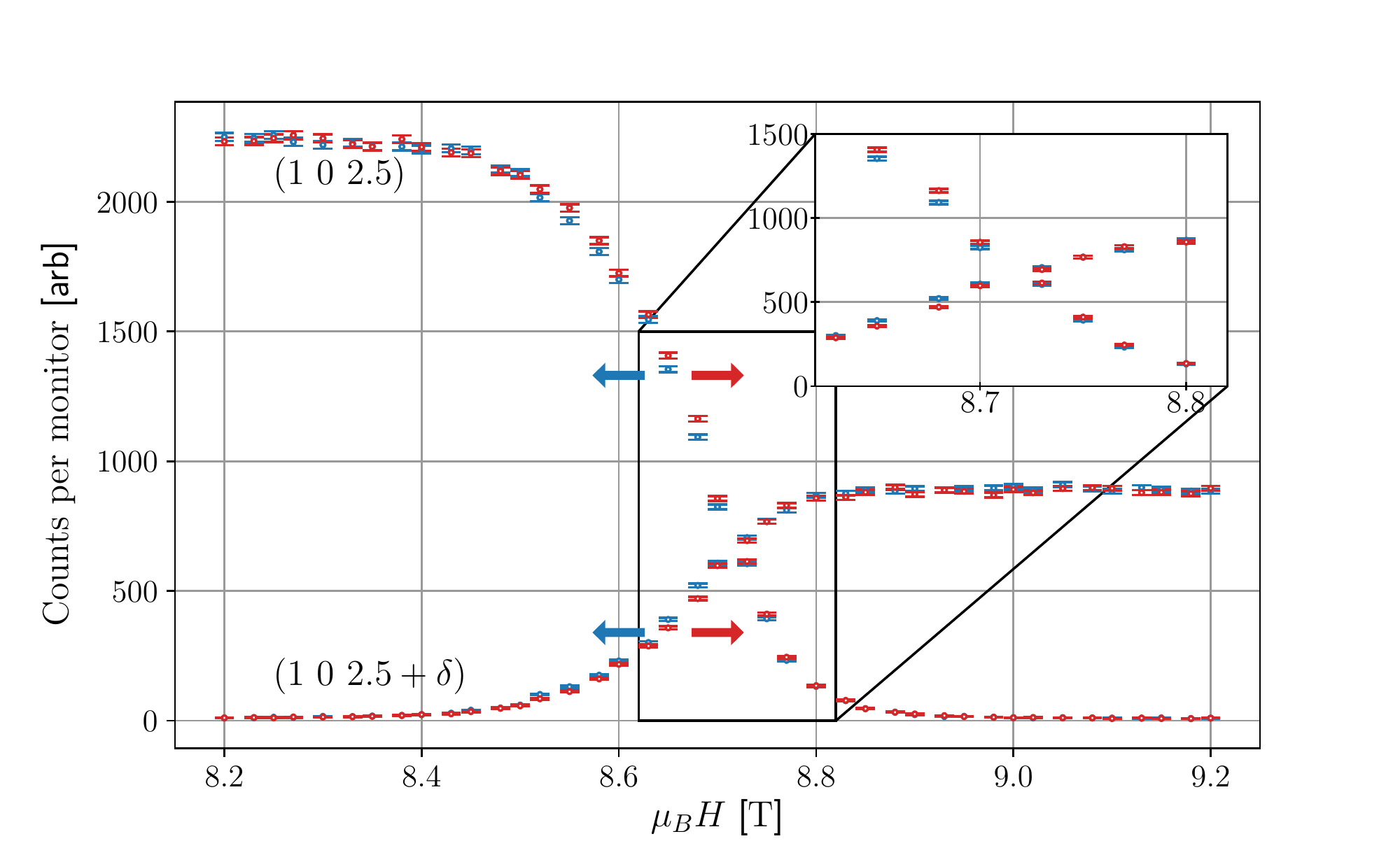}
\caption{(top) Phase diagram of NTO in the $(B,T)$ plane with $B||c$, showing the commensurate (C), incommensurate (IC), co-existence (CE), and paramagnetic (PM) phases. Dotted lines are guides to the eye. Data was taken at RITA-II using the $(1 \; 0 \; 2.5)$ and $(1 \; 0 \; 2.5+\delta)$ reflections. (bottom) Magnetic hysteresis measured by diffraction intensity at the incommensurate $(1 \; 0 \; 2.7)$ peak taken at RITA-II at 1.7~K denoted by arrows in the phase diagram, shown as a function of field during field ramps. 
} 
\label{fig:phasediagram}
\end{figure}

The very clear result is that the spins obtain an incommensurate in-plane component with modulation vector of $\delta \sim 0.18$ along the $c$-axis. Fig.~\ref{fig:rawdataANDmagnetwindows} additionally shows that this modulation vector jumps discontinuously from commensurate to incommensurate upon increasing field, with a region of co-existence of $\approx$ 0.4~T. 
The exact extent of the co-existence area might depend on the accuracy of the field alignment to the sample $c$ axis~\cite{Callen65} but only a first order transition can provide the observed discontinuous jump of the ordering vector. 
Investigating the co-existence region further, we performed hysteresis scans by monitoring the intensity of the $(1 \; 0 \; 2.5+\delta)$ peak, while ramping magnetic field through the phase transition at $T=1.7$~K, c.f. bottom of Fig.~\ref{fig:phasediagram}. The field ramp was halted for 10~s before each measurement to ensure thermal equilibrium. 
From this data, we conclude that there is no evidence for hysteresis in agreement with current literature~\cite{Oh2014}.
\begin{figure}[ht]\centering
    \centering
    \includegraphics[width=\linewidth]{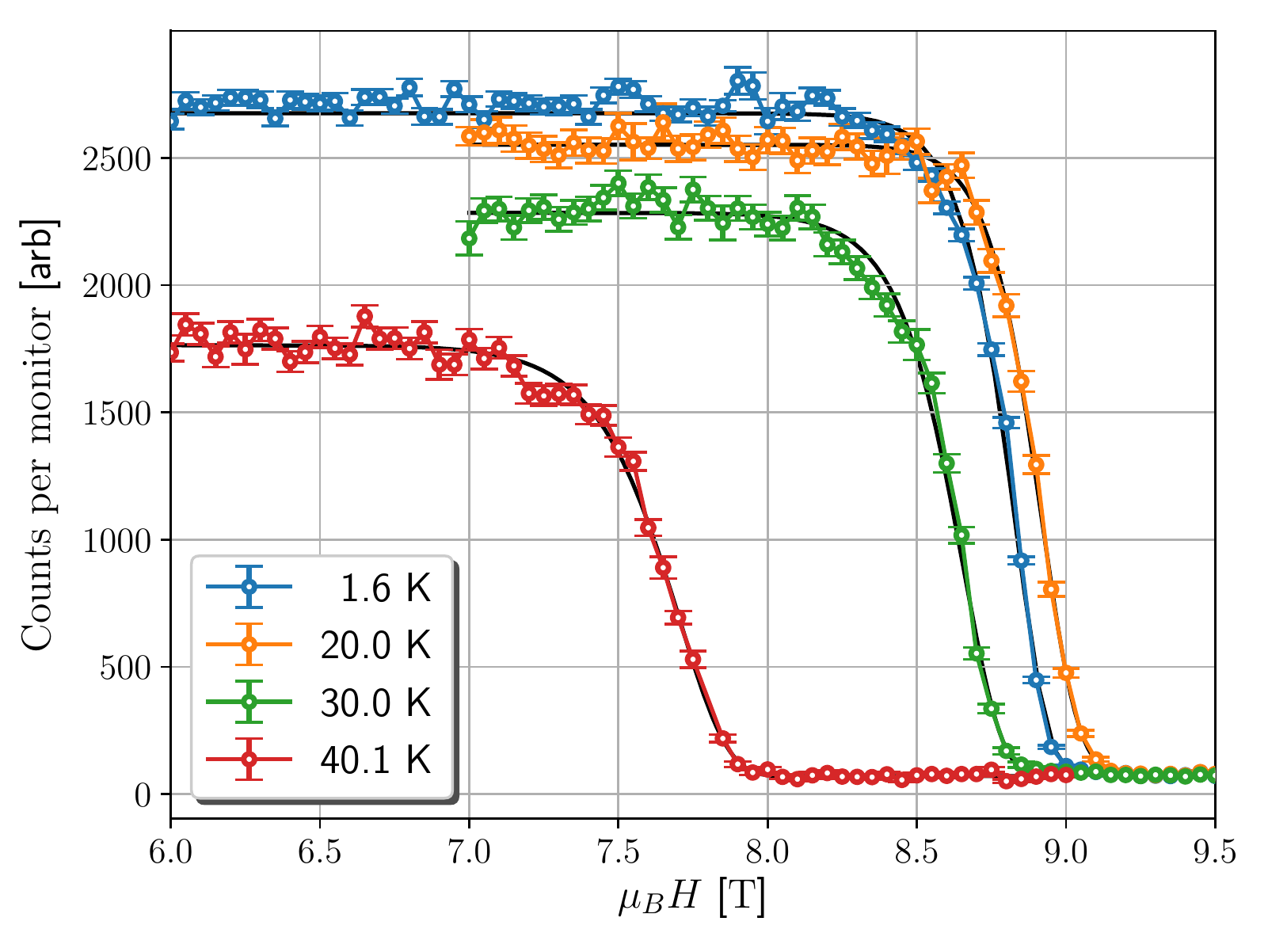}
    \caption{Intensity of the commensurate $(1 \; 0 \; 2.5)$ measured at RITA-II across phase transition from C to IC at different temperatures. Solid lines are fit to the data.}
    \label{fig:criticalB}
\end{figure}
From investigations of the phase transition at a number of temperatures, we are able to produce a $(B,T)$ phase diagram of the ordered spin states, the coexistence (CE) and paramagnetic (PM) phases, shown in Fig.~\ref{fig:phasediagram}. A similar phase diagram has already been measured using the dielectric constant\cite{Yokosuk2016}, but our diagram has the additional feature of defining the co-existence phase. As the transition is of first order, an exact mathematical description of the ordering parameter is not present. It was found that performing a fit using the Sigmoid function $I(T)=a/(1+b^{T-T_n})+B$ gave the most reliable results across all temperatures, c.f. Fig.~\ref{fig:criticalB}. Then, the CE phase is defined as the region between 90\% and 10\% of the full intensity for the commensurate peak. Futher, our results agree with the PM to C transition temperature, $T_{\rm N}=52$~K in zero field. 
For the second-order thermal phase transitions between PM and C, and PM and IC, the critical exponents are $\beta_C = 0.298\pm0.005$, and $\beta_{IC} = 0.353\pm0.011$ respectively, see Fig.~\ref{fig:criticalT} in the appendix.

\section{Magnetic Structure Determination}
From the observation of pairs of incommensurate peaks we deduce the existence of an incommensurate high-field magnetic structure. This structure is at variance with the interpretation of the earlier bulk magnetization/susceptibility studies, which concluded a continuous spin-flop transition, maintaining the commensurate ordering vector~\cite{Oh2014,Yokosuk2015}.
\begin{table}[h]\centering
\begin{tabular}{c|c|c|c}
H [rlu] & K [rlu] & L$-\delta$ [rlu] & L$+\delta$ [rlu] \\ \hline

-4 & 0 & 12.649 $\pm$ 0.006 & --- \\ 
-2 & 0 & 4.377 $\pm$ 0.005 & --- \\ 
-2 & 0 & 5.3226 $\pm$ 0.0006 & 5.6945 $\pm$ 0.0006\\ 

-2 & 0 & 8.3213 $\pm$ 0.0014 & 8.705 $\pm$ 0.003 \\ 

0 & 0 & 1.33533 $\pm$ 0.00011 & 1.69780 $\pm$ 0.00011\\ 
1 & 0 & --- & 2.68453 $\pm$ 0.00012$^*$ \\
1 & 1 & --- & 4.6928 $\pm$ 0.0005\\ 
1 & 1 & --- & 7.671 $\pm$ 0.005\\ 
2 & 0 & 4.415 $\pm$ 0.009 & --- \\ 
2 & 0 & 6.316 $\pm$ 0.006 & --- \\ 
\hline
\multicolumn{2}{c|}{$\delta_{\pm}$} & 0.16511$\pm$0.00011 & 0.19738$\pm$0.00011 \\ \hline

\end{tabular}
\caption{Positions of measured incommensurate peaks at field of 9.9~T and T $\sim 1.8$~K. Data taken at EIGER. The incommensurate peak at $(2 \;  0 \; 6.7)$ is not measured due to instrumental limitations. $^*$ Point not included in averages as it was taken at 9.5 T and 25 K. } \label{tab:magneticreflections}
\end{table}


While the geometrical restrictions of the horizontal-field magnet prevented us from obtaining a data set of a quality that can be used for detailed magnetic structure refinement, 
we have sufficient data to perform a representational analysis. The result is that for the given ordering vector the only allowed symmetry is an incommensurate circular spiral together with a ferromagnetic component, i.e. the conical spiral with ordering vector $(0\;0\;1.5+\delta)$. Due to a small misalignment in our diffraction experiment two slightly different values of $\delta$ are found, Tab.~\ref{tab:magneticreflections}, of the size 0.015 $l$ corresponding to an offset in A4 of 0.15$^\circ$, which is an acceptable misalignment on a triple axis instrument. The conclusion of a cross-over with spins flipping into the a-b plane and incommensurable $\delta$ is reached by noting that the presence of peaks along (0 \; 0 \; $l$) requires a spin component in the a-b plane and that the observed continuously shifting position of $\delta$ with $B$ or $T$ requires an incommensurate structure. 
From bulk magnetization measurements~\cite{Kim2015} it is known that the magnetization in the high field phase increases with field strength. This is not possible in the incommensurate rotating phase described by the irreducible representations $\Gamma_2\oplus\Gamma_3$, Tab.~\ref{tab:IrreducibleRepresentations}, as the total magnetization is zero. It can then only be explained by a ferromagnetic component along $c$, described by $\Gamma_1$. In total, the only possibility is a linear combination of $\Gamma_1$ and $\Gamma_2\oplus\Gamma_3$. The relative phases and tilts remain to be found from a magnetic structure refinement. Combined with the ordering vector along (0 \; 0 \; $l$) a spin spiral is found to be the unique solution~\cite{Rodriguez-Carvajal2012}.

\begin{table}[ht]\centering
\begin{tabular}{c|c|c|c}
 & $\Gamma_1$ & $\Gamma_2$ & $\Gamma_3$ \\ \hline
Ni$_n$ & $\begin{pmatrix}0\\0\\u\end{pmatrix}$ &  $\begin{pmatrix}\frac{3-i\sqrt{3}}{2}u\\-i\sqrt{3}u\\0\end{pmatrix}$ & $\begin{pmatrix}\frac{3+i\sqrt{3}}{2}u\\i\sqrt{3}u\\0\end{pmatrix}$\\ \hline
\end{tabular}
\caption{Irreducible representations of the $n$'th magnetic Nickel atom in the unit cell.} \label{tab:IrreducibleRepresentations}
\end{table}

\section{Origin of the ferro-electric transition}
As the electric polarization and the ordering vector are parallel, the DM-interaction is excluded as origin of the transition, while the symmetric exchange allow this~\cite{Tokura2014}. This symmetric exchange fits qualitatively with the observed decrease in electric polarization across the phase transition\citep{Kim2015}, as shown by Landau theory\cite{Oh2014}, where the spins rotate away from the AFM structure, which has the highest polarization, into the conical spiral.


How quickly the spins rotate around the $c$ axis, i.e. the value of the incommensurability $\delta$, depends strongly on the detailed balance of the spin couplings. Changing the distance or angle between two Ni atoms would change their coupling and thus in turn change $\delta$. From studying the ($q=0$) optical phonon frequencies both across the magnetic field transition as well as function of temperature, it has been found that the phonon modes at $f=310$ cm$^{-1}$, $f=597$ cm$^{-1}$, and $f=666$ cm$^{-1}$ behave unexpectedly~\cite{Yokosuk2015}. Across the magnetic phase transition these modes are perturbed while other modes are unchanged. This points towards a M-E coupling that acts to modify the Ni-atom position depending on spin direction and should thus be observable in the IC phase by changing either magnetic field or temperature. This effect is found in the elastic data shown in the top of Fig.~\ref{fig:rawdataANDmagnetwindows}, where the center of the IC peak can be seen to move to slightly lower $l$-values when comparing its onset with its stable position at B$>$8.7~T. 

\begin{figure}[ht]\centering
\includegraphics[width=\linewidth]{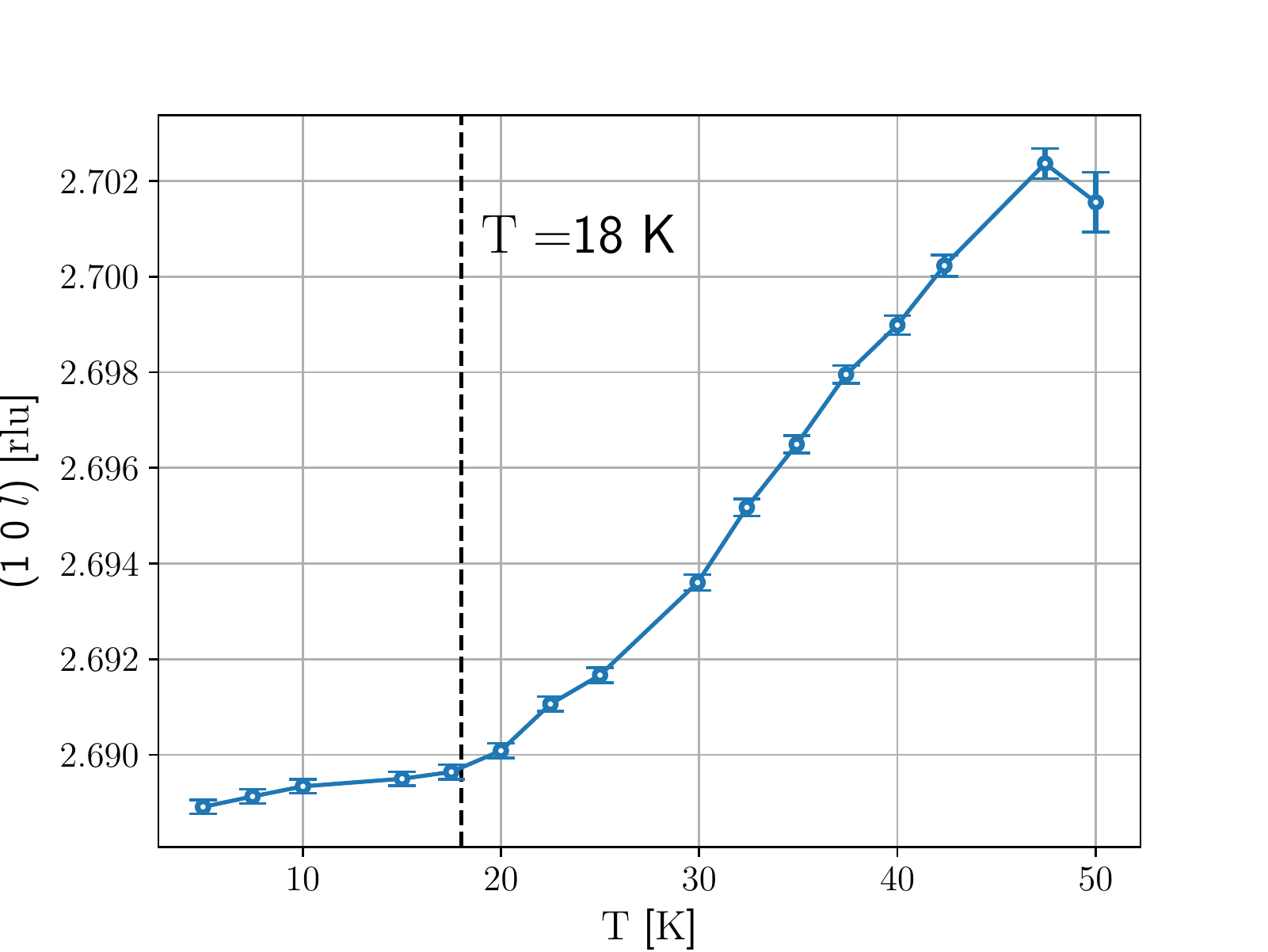}
\caption{Temperature variation of the magnetic peaks position at $(1 \; 0 \; l)$, measured at $B=9.9~T$ at RITA-II.} 
\label{fig:Lvalue}
\end{figure}

\begin{figure}[ht]\centering
    \includegraphics[width=\linewidth]{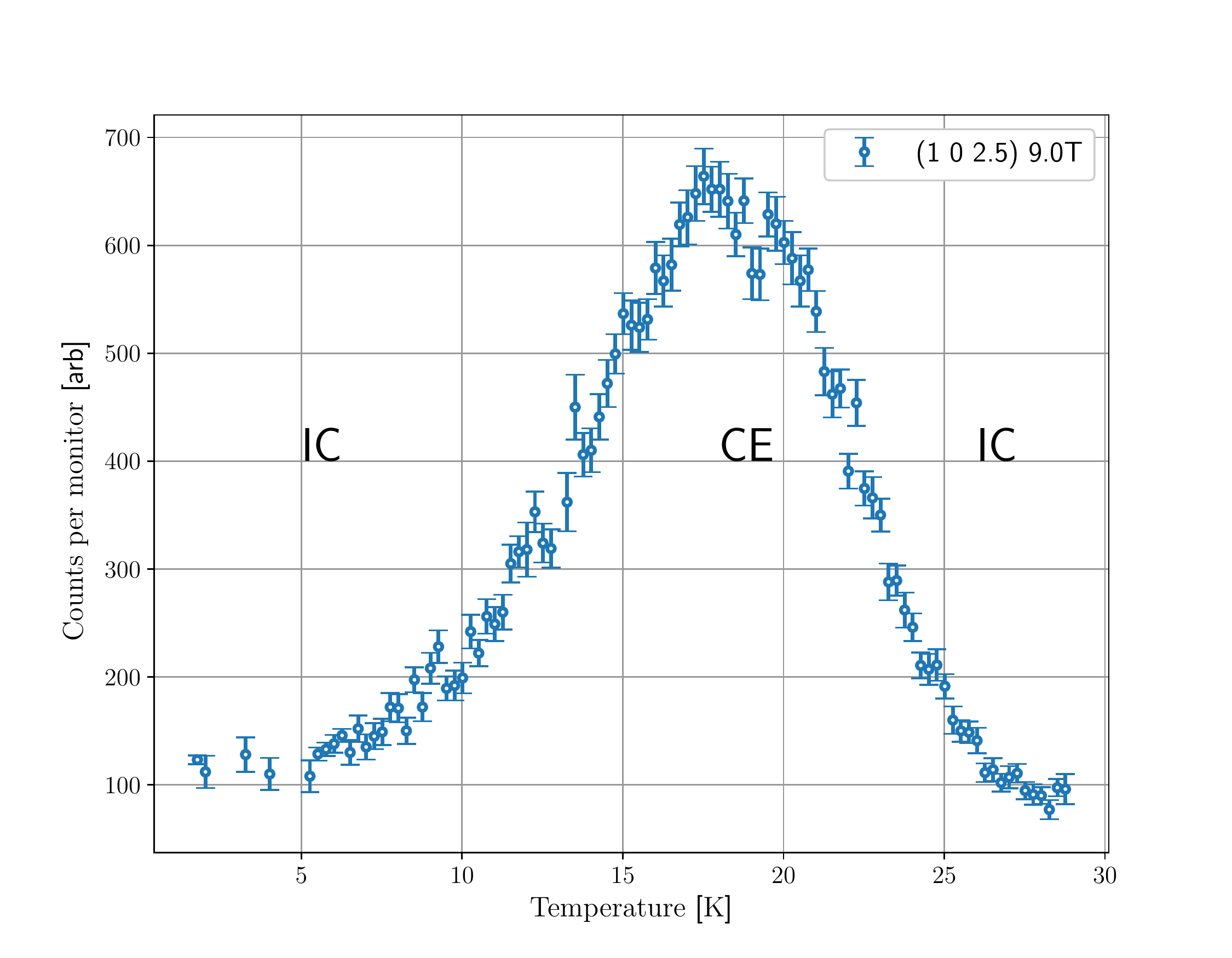}
    \caption{Intensity of the commensurate peak at 9.0 T as a function of temperature across the re-entry area. As seen, the system starts out at low temperature in the IC phase, goes into the CE phase upon heating, while ending at 27 K in the IC phase again.}
    \label{fig:re-entry}
\end{figure}

A closer look at the IC position as a function of the temperature reveals that its position also varies with temperature, c.f. Fig.~\ref{fig:Lvalue}. It is seen that the peak position moves slowly towards larger $l$-values for temperatures below approx. 18~K above which it moves more quickly. A temperature effect is also observed through the phase boundary at constant magnetic field and changing temperature where a re-entrant behaviour is found for a constant magnetic field of 9.0~T, c.f. Fig.~\ref{fig:re-entry}.


\section{Simplified magnetic Hamiltonian}
In this section, the most simple magnetic Hamiltonian capturing the effective behaviour of NTO at low energy and low magnetic field is presented. To simplify this system the most, a unit cell consisting of only three magnetic atoms is used as opposed to the 9 found in NTO. Further, the spins are assumed to be placed equidistant from each other along the c axis. In effect, this means that there will be no distinction between the different Ni atoms in the simplified Hamiltonian, further leading to the low-field structure changing from the $\uparrow\uparrow\downarrow\downarrow\downarrow\uparrow$ to the simpler $\uparrow\downarrow\uparrow\downarrow$ sequence, the impact of which will be discussed later. 

\begin{figure}[ht]\centering
\includegraphics[height=0.7\linewidth]{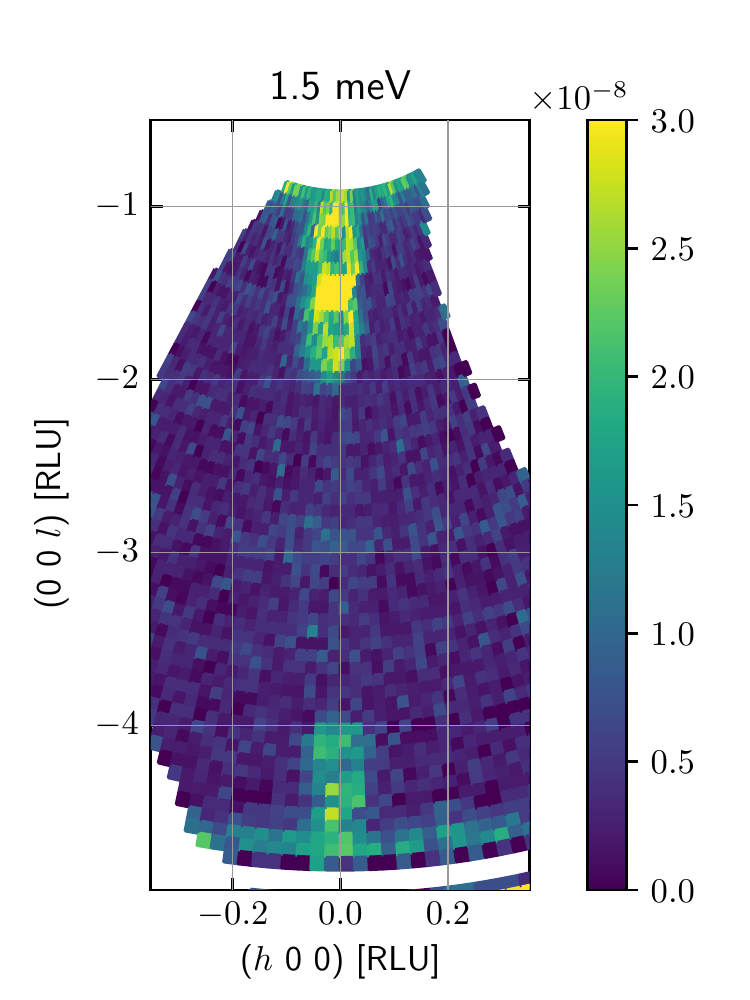}
\includegraphics[height=0.65\linewidth]{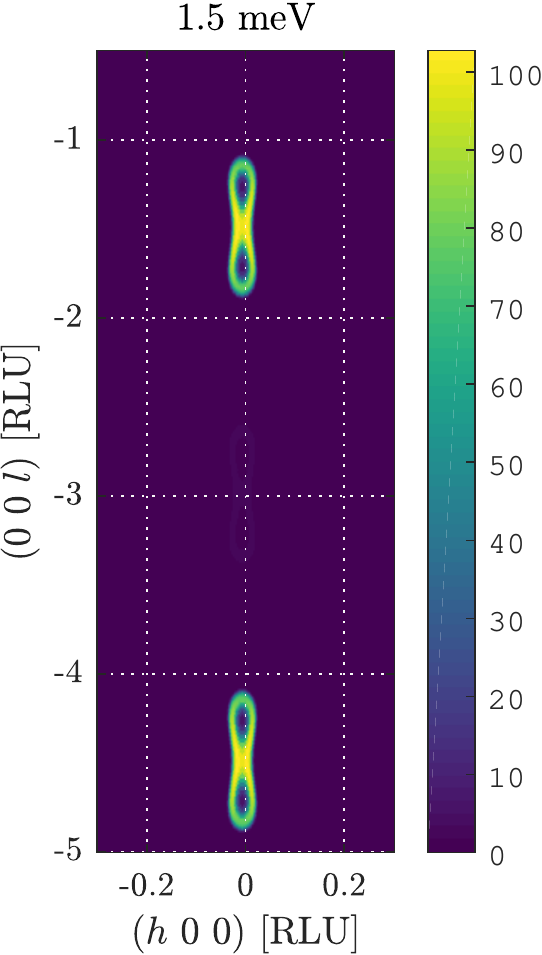}
\caption{(left) Map of inelastic scattering intensity from NTO taken at an energy transfer of $\Delta E$ = 1.5 meV, and showing a large part of the ($h\; 0 \; l$)-plane. Data taken at MultiFLEXX. (right) Simulated excitations from SpinW as described in the main text. The maps show the bottom of the dispersion relation, which is placed exactly at $Q_{\rm gap} = Q_{\rm IC}$.} 
\label{fig:inelastic} 
\end{figure}

\begin{figure}[!htb]
    \centering
    \includegraphics[width=0.95\linewidth]{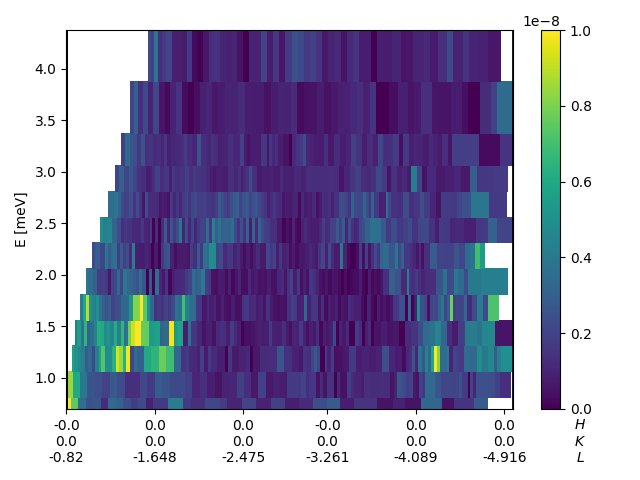}
    \includegraphics[width=0.95\linewidth]{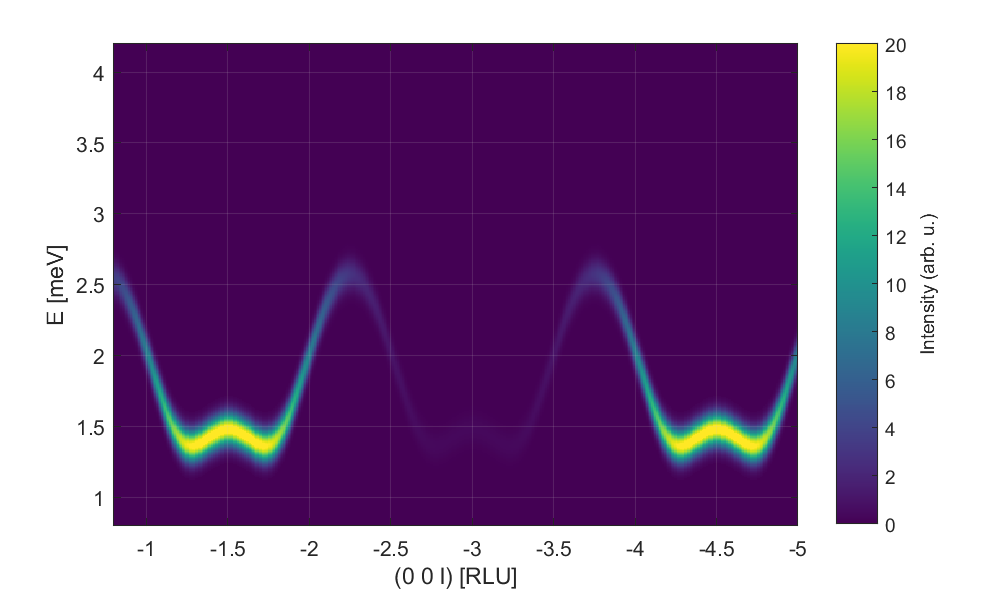}
    \caption{
    (top) Intensity as function of energy transfer and $Q$ along $(0\; 0\; l)$ measured at MultiFLEXX in zero magnetic field and 2K. (bottom) Excitation spectrum for simplified magnetic Hamiltonian optimized to mimic NTO.}
    \label{fig:InelasticVertical}
\end{figure}

The inelastic neutron scattering experiment was performed using a standard Orange He-flow cryostat. Utilized the CAMEA-type MultiFLEXX back-end. 
The data taken consisted of rotating the sample by 180$^\circ$ in steps of 1$^\circ$. For each sample orientation 4 values of the scattering angle was used to cover dark angles and extend coverage. Also using multiple settings of the initial neutron energy gave the data in shown Figs.~\ref{fig:inelastic} and \ref{fig:InelasticVertical}.

The qualitative behaviour sought for is first of all the incommensurate low energy spin excitation with minimum in $Q_{\rm IC}$, an energy gap of around 1.2 meV, a critical magnetic field of 8.6 T, and the overall extend of the magnon as shown in our inelastic data. With an effective moment of 2.03(2) $\mu_B$ or spin 1\cite{Zivkovic2010} and a spin wave gap of 1.2 meV, the critical magnetic field is expected to be around 10.3 T, which is closet to what is observed. For the excitation to have an incommensurate minimum, the excited spin structure has to be able to rotate; in this case around the $c$ axis. The simplest way that allows this is if the spins gain a component in the a-b plane. This requires a spin canting, which can be enforced by the rudimentary spin flop model relying only on an AFM coupling, $J_1$, and a small uniaxial anisotropy along $c$, $\Delta$, which is included to ensure that the spins are collinear in low fields, and then a coupling to a magnetic field\cite{Blundell}. 

Next, the incommensurability is known from the fact that the magnon has minimum at the $Q_{\rm IC}$ position is $(0\; 0\; 1.5\pm\delta)$ for $\delta = 0.18$. This corresponds to a rotation of all spins by 1.32 or 1.68 per unit cell, or equivalently 158.4$^\circ$ or 201.6$^\circ$. To achieve this helical behaviour an antiferromagnetic term is to be added to the Hamiltonian. By adding a $J_2$ between next nearest neighbours along the $c$ axis with a strength of $J_2=-\frac{|J_1|}{4\cos{\theta}}$ it is ensured that the minimum for rotation around the $c$ axis is at $\theta$\cite{Blundell}. 

Lastly, the spin wave is known to disperse rather sharply in the $a$ direction while being more elongated along the $c$ axis giving an impression of the relative strengths of couplings in these directions. Further, the ordering is ferromagnetic in the a-b plane giving the positive sign of the coupling in this direction $J_{\rm FM}$. Collecting everything, the effective magnetic Hamiltonian becomes

\begin{align}
    H &= J_1\sum_{nn ,c}\vec{S}_i\cdot\vec{S}_j+J_2\sum_{nnn, c}\vec{S}_i\cdot\vec{S}_j+\Delta\sum_iS_{i,z}^2 \notag\\ 
    &+\sum_iS_{i,z}g\mu_BH_z+J_{FM}\sum_{nn, ab}\vec{S}_i\cdot\vec{S}_j, \label{eq:SimplifiedHamiltonian}
\end{align}
where nn and nnn denotes the nearest and next-nearest neighbour respectively. 
The values found to best mimic NTO when comparing to the observations above are tabulated in Table~\ref{tab:SimplifiedHamiltonian}. 
\begin{table}[ht]
    \centering
    \begin{tabular}{c|c|c|c|c} 
        Parameter &  $J_1$ & $J_2$ & $J_{FM}$ & $\Delta$ \\\hline 
        Value [meV] & 2.55 & 0.6856 & -2.75 & -0.1045 \\\hline 
    \end{tabular}
    \caption{Coupling strengths used in the simplified Hamiltonian in eq.~(\ref{eq:SimplifiedHamiltonian}).}
    \label{tab:SimplifiedHamiltonian}
\end{table}
\begin{figure}[!htb]
    \centering
    \includegraphics[width=0.45\linewidth]{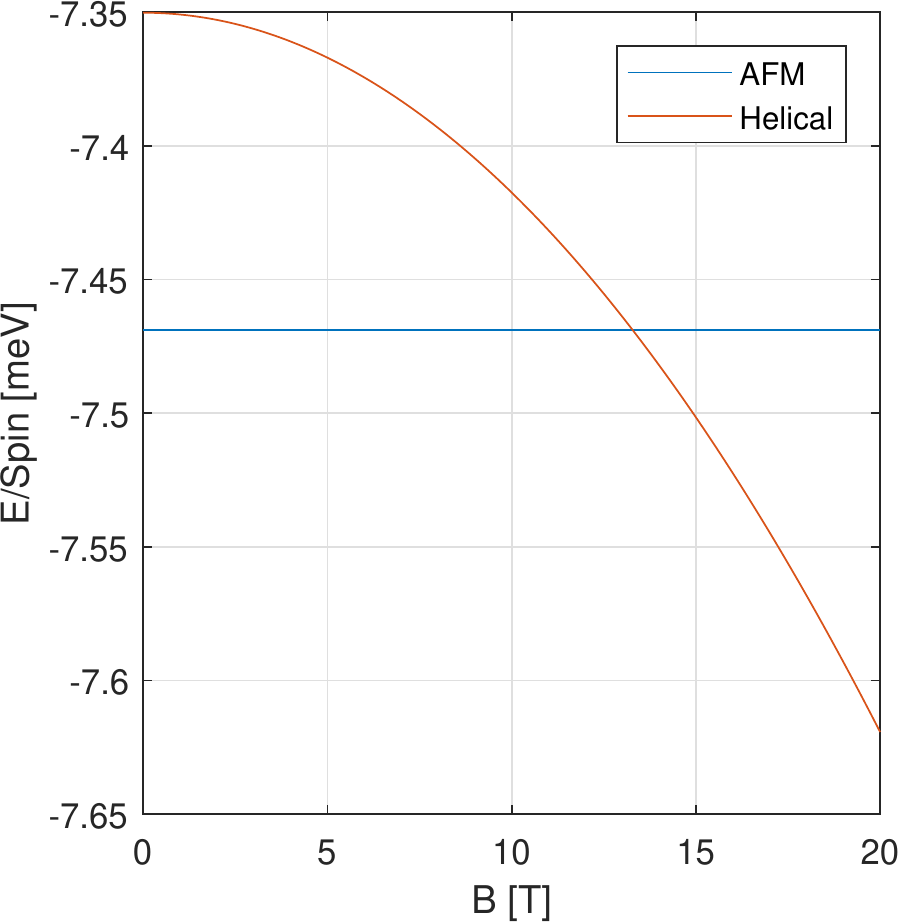}
    \includegraphics[width=0.45\linewidth]{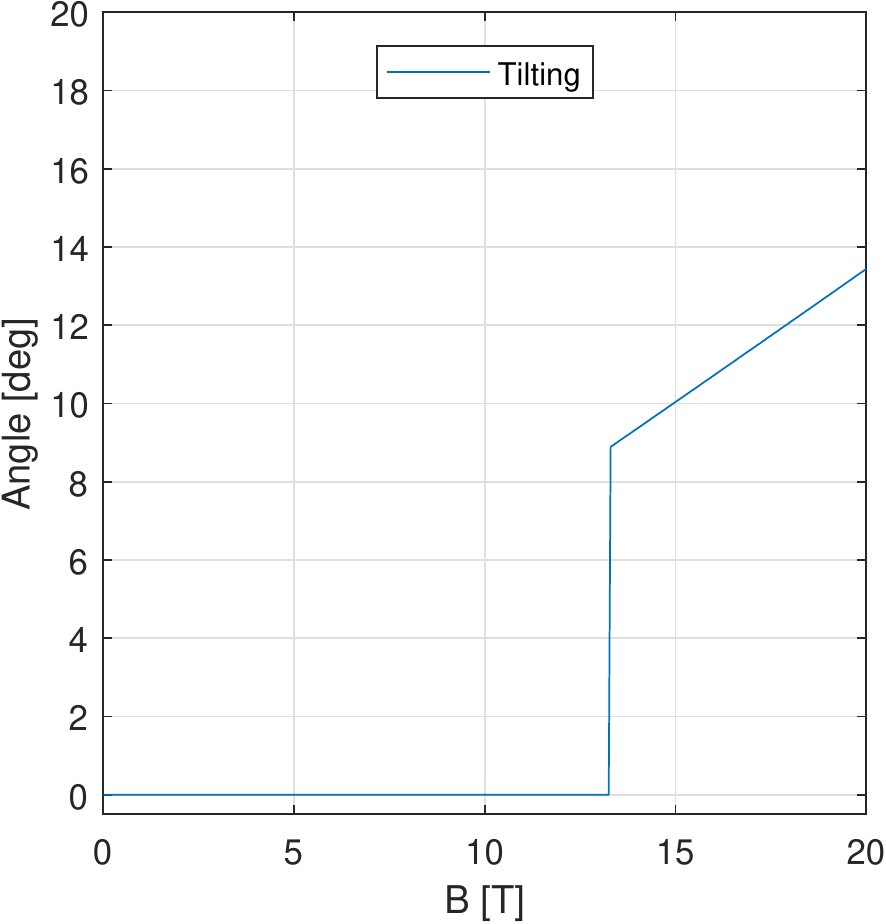}
    \caption{
    (left) Energy of AFM low field ground state and helical state as function of magnetic field. (right) Tilting angle of spins with respect to the $c$ axis as function of magnetic field.}
    \label{fig:SimpleHamiltonAppendix}
\end{figure}
These were found, starting from the $J_{\rm FM}$ being $-2.75$ meV and $J_1$ 2.55 meV as these gave the desired steepness in the spin wave spectrum both along and perpendicular to $c$. Next, to ensure the incommensurability $J_2$ has to be 0.6856 meV, and for the spin wave gap to be comparable to that in Fig.~\ref{fig:InelasticVertical} it is to have a value of around -0.1045 meV.

These values ensure that the low E behaviour of the model resembles NTO. 
More specifically, the IC spin waves and their gap is found together with a similar maximal magnon energy around the $(0\; 0\; 2.5)$ position; comparing 2.7 meV from NTO to the 2.6 from the model. Comparing the steepness of the magnon dispersion the model is more steep than the experiment as seen from the distance between the maximum points, c.f. Fig.~\ref{fig:InelasticVertical}. Lastly, the experimental data show the spin excitation to actually consist of two branches whereas the model only has one. This means that the NTO magnons with minima at $(0\; 0\; 1.5\pm\delta)$ are to be modelled as two separate modes.


Regarding the phase transition, the critical field of the model is somewhat higher than for NTO, namely 13.28 T as compared to the $\sim$ 8.6 T. This most likely stems from the simplification of the low field structure to only consist of 3 Ni atoms and that their spin structure is fully antiferromagnetic while that of NTO is the more complex $\uparrow\uparrow\downarrow\downarrow\downarrow\uparrow$ that may lead to some couplings being frustrated. In effect, this would reduce the energy difference between the low and high field phases resulting in a lowering of the critical magnetic field. However, our model predicts the phase transition to occur simply via a first order transition between the AFM and helical structures simply due to the latter becoming more energetically favourable at higher magnetic field. What happens is that the spins are locked to be along the $c$ axis as long as the system orders AFM but at the phase transition the spins flop out into the a-b plane. Their component along the $c$ axis is now allowed to change and they slowly tilt to align along the $c$ for increasing magnetic field strengths. This tilting is seen in Fig.~\ref{fig:SimpleHamiltonAppendix} where its value is found from minimizing the total energy of the system with respect to tilting; only after the phase transition has appeared.

Investigating our simple model further at $B_C$ we find the energy landscape to be so shallow along $l$ (0.02 meV/spin) at the phase transition that thermal fluctuations even at 2~K may cause the system to cross the phase boundary locally, giving rise to the absence of hysteresis observed. We speculate that a neutron diffraction or magnetisation experiments performed at milllikelvin temperatures could show a measurable hysteresis.

In summary, the simple Hamiltonian in eq.~\eqref{eq:SimplifiedHamiltonian} captures many of the measured features of NTO. However, the simplification of the unit cell and its low field spin structure impacts the spin excitations by only allowing a single excitation in stead of the two seen. Further, the exact field value for the transition is off by $\sim$ 50\%. Once more, this is due to the simplifications made, and the fact that the model does not incorporate any exchange striction and magneto elastic terms. A full model of NTO should incorporate these together with the full unit cell.

\section{Conclusion}
We find that the magnetic moments in Ni$_3$TeO$_6$ change from a commensurate collinear antiferromagnetic structure with spins along $c$ at low fields and temperature to a conical spin spiral with propagation vector $q_{\rm IC} \approx (0 \; 0 \; 1.5 \pm \delta)$; $\delta \sim 0.18$ at high fields along $c$. We observe a large region of co-existence, 0.4~T wide, between the two phases with negligible hysteresis. From our representational analysis we find that the magnetic structure is a conical spiral and that the inverse DM-effect is excluded as the driving force for the multiferroicity.

The combined evidence points to a field-driven first order C-IC transition between the two phases. 
The exact value of the incommensurability is temperature and magnetic field dependent, attributed to the fine balance needed between coupling constants in part determined by the movement of the magnetic Ni-ions. 
A simple model of the magnetic Hamiltonian is able to reproduce our findings qualitatively and show that the energy landscape at the phase transition is shallow enough for thermal excitations at 2~K to explain the lack of hysteresis.

\acknowledgments
We are indebted to Jose Antonio Alonso for support and access to his synthesis laboratory. We thank Jonathan White for creating the software for planning experiments in the restricted geometry of the horizontal-field magnet at SINQ. The project was funded by the Danish Agency for Research and Innovation through DANSCATT. JL was supported by Nordforsk through NNSP, and by the Paul Scherrer Institute. We also thank Simon Ward for assistance with the SpinW computations.

Neutron scattering experiments were performed at the SINQ neutron source, Paul Scherrer Institute, Villigen, Switzerland and at the BER-II facility, Helmholz Center, Berlin.

\bibliography{Ni3TeO6bib}
\clearpage
\appendix
\newpage
\section*{Appendix}
\FloatBarrier

\subsection{Order of the phase transition}
\begin{figure}[ht]\centering
\includegraphics[width=\linewidth]{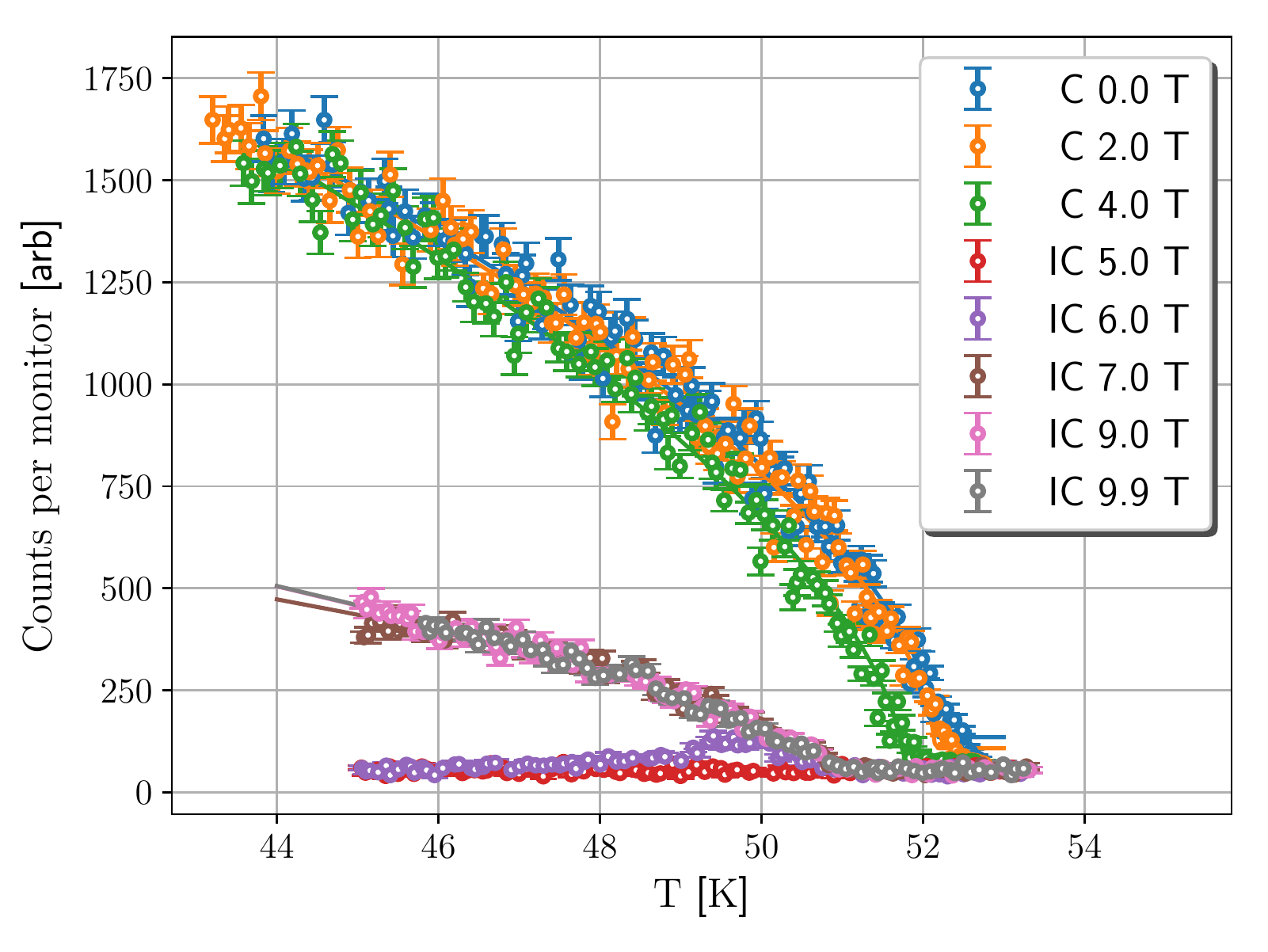}
\caption{Temperature variation of the C and IC peak intensities at different fields across the phase transition from the PM phase.} 
\label{fig:criticalT}
\end{figure}

To determine the order of the phase transition, we investigated the critical behavior of the magnetic intensity along the phase boundaries between the three phases. The data is shown in Figs.~\ref{fig:criticalT} and \ref{fig:criticalB}. We see that the temperature dependence of the peak intensities, corresponding to the phase boundary to the PM phase, can be well described by a power law, $I(T) \propto t^{2\beta}$, with the reduced critical temperature being $t = (T_{\rm N}-T)/T_{\rm N}$. This is a defining feature of a second-order phase transition. The critical exponent is found to be in the range $\beta = 0.25 - 0.38 $. 

In contrast, the field dependence of the magnetic intensities between the C and the IC phases can clearly not be described by the power law behaviour, c.f. Fig.~\ref{fig:criticalB}. In addition, the peak position moves discontinuously between the two phases without any critical scattering, c.f. Fig.~\ref{fig:rawdataANDmagnetwindows}. Therefore, we have strong evidence to say that the C-to-IC phase boundary is of first order.



\subsection{Representational analysis} 
From the Fourier projection of the spin direction into the irreducible representation for NTO one sees that three principal axis are present, $\Gamma_1$, $\Gamma_2$, and $\Gamma_3$, listed in Tab.~\ref{tab:IrreducibleRepresentations}. Here $\Gamma_1$ describes the commensurate, low-field structure, while $\Gamma_2$ and $\Gamma_3$ describe spiral structures and are connected by complex conjugation. A spin structure containing one of $(\Gamma_2, \Gamma_3)$ would usually also contain the other as to ensure the spins be real. The reasoning behind the combination of $\Gamma_2$ and $\Gamma_3$ is to ensure the magnetic moment to be real. Notice that $\Gamma_2 = \Gamma_3^*$, and that $\Gamma_2\perp\Gamma_3$, as they are described in the hexagonal coordinate system, and their lengths are equal. This combination then represents a circular rotations of the spins in the a-b plane.

All of the representations are found from the R3 symmetry of the space group, 146, thus a transition between $\Gamma_1$ and $(\Gamma_2, \Gamma_3)$ does not break any crystal symmetry. 

As for the transition from the high-temperature paramagnetic phase to the commensurate low-temperature phase, an AFM order is created suggesting a second order phase transition. The same is the case for this transition at high field where the paramagnetic phase is transformed into the spiral phase. However, between these two phases the magnetic order neither disappears nor does a crystal symmetry break. Thus, a transition not being of second order is fully allowed. This is believed to be the case where the magnetic ordering vector abruptly changes from being (0 \; 0 \; 1.5) to (0 \; 0 \; 1.5$\pm \delta$) and thus tips the spin components from being only along $\Gamma_1$ to also be along $\Gamma_2\pm\Gamma_3$, where both left and right handed rotation is possible. 

This in-plane rotation is also consistent with the presence of scattering intensity at (0 \; 0 \; l) peaks as well as the drop in electric polarization as described in the main text.

\end{document}